**Ultrahigh electrostrain in Pb-free piezoceramics: Effect of bending**

Gobinda Das Adhikary[1#], John Daniels[2], Luke Giles[2], Rajeev Ranjan[1*]

Department of Materials Engineering, Indian Institute of Science, Bengaluru-560012, India.

[2]School of Materials Science and Engineering, UNSW Sydney, Kensington, Sydney, Australia, 2052

## Abstract

Recently several reports showing ultra-high electrostrain (> 1 %) have appeared in Pb-free piezoceramics. However, there is lack of clarity on the nature of the ultrahigh strain. Here, we demonsrate that the ultrahigh strain is a consequence of bending of the disc. We show that the propensity for bending arises from the difference in the response magnitude of the grains at the positive and negative surfaces of the piezoceramic when the field is applied.

[#] gadhikary53@gmail.com

*rajeev@iisc.ac.in

Recently a series of publications showing large electrostrain > 1 % in different lead-free piezoceramics [1 - 7]. An implicit assumption in all reports claiming the exceptionally large electrostrain > 1 % is that the electrostrain is longitudinal, i.e., the disc's thickness increases on the field application along the thickness direction. Moreover, this has been attributed primarily to the introduction of defect dipoles in the system via non-stoichiometric chemical substitutions. Different groups have envisaged different chemical modification strategies to create defect dipoles to achieve ultrahigh strain [1-7] . In contrast to these claims, Adhikary and Ranjan [8], Adhikary et al [9] have shown that thick discs (0.7 mm) of the same compositions show low strain ~ 0.2 -0.3 %. Only when the thickness was reduced to ~ 0.2 mm, the measured strain values could match or exceed the reported ultrahigh strain. For example, Luo et al have reported 1.1 % strain at 100 kV/cm in an oxygen-deficient Pb-free composition (1-x)$Na_{0.5}Bi_{0.5}TiO_3$-(x)$BaAlO_{2.5}$ [5]. Tina *et al.* [10] reported that thicker piezoceramic discs (0.7 mm) of the same piezoelectric system show a maximum unipolar strain of 0.35 %. When the thickness is reduced to 0.15 mm, the maximum unipolar strain reaches ~ 5.5 % [10]. The bipolar strain shows 10 % in the positive cycle and nearly zero value in the negative cycle [9, 10]. Since such anomalous strain values cannot be rationalized in terms of switching of ferroelectric-ferroelastic domains, field-driven phase transformation, and lattice strain, it is important to resolve the nature of ultrahigh strain and the associated mechanism. In this paper, we resolve this issue and show that the ultrahigh strain is a consequence of the bending of the disc on application of electric field due to asymmetric response of the grains near the positive and the negative potential of the applied field.

We performed experiments on (1-x)$Na_{0.5}Bi_{0.5}TiO_3$-(x)$SrTiO_3$ with x=0.26, a composition at the crossover of ergodic and non ergodic relaxor ferroelectric state [11]. The specimens were prepared using the conventional solid state sintering method, the details of which can be found in ref 11. Electrostrain and polarization measurements were carried out on ~ 95 % dense sintered discs with Radiant ferroelectric measurement system (model Premier Precision II) Measurements were performed on circular discs (diameter 10-12 mm) of different thicknesses in the thickness regime 1 mm – 0.2 mm.

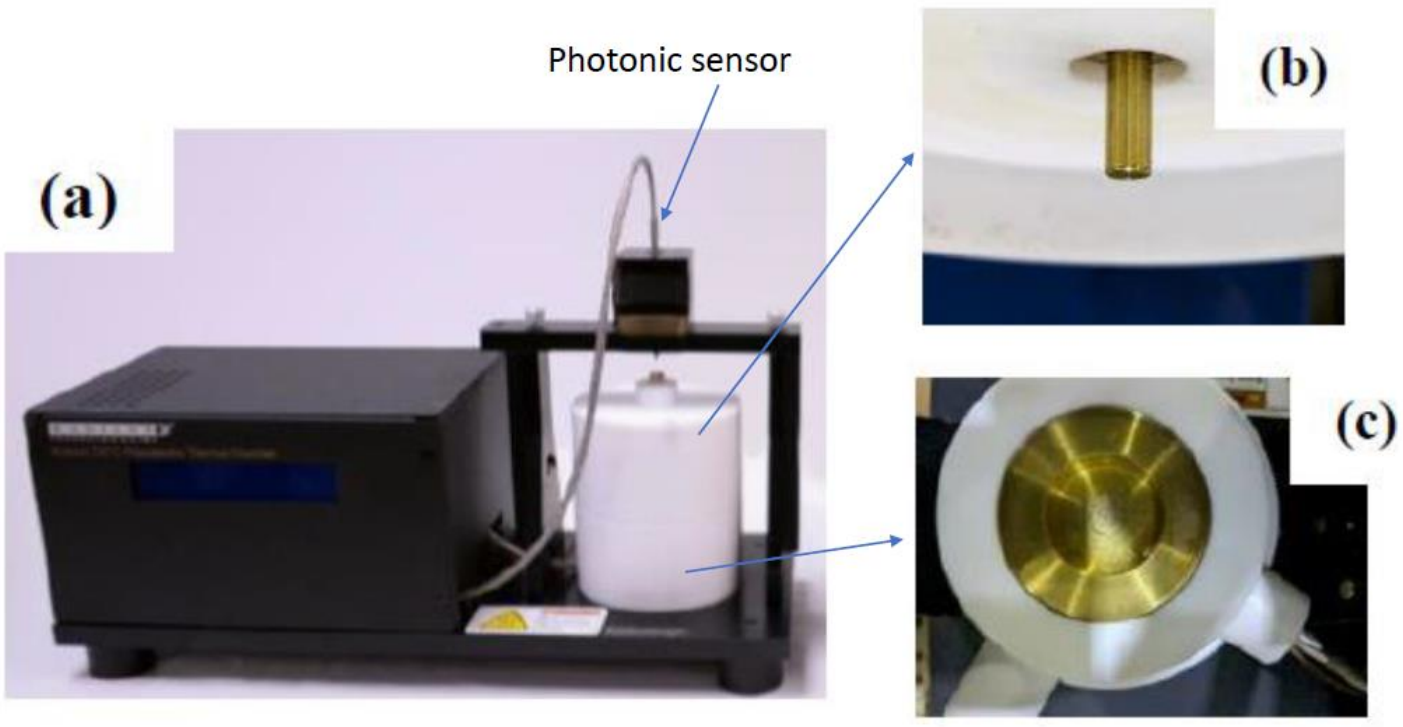

Fig. 1 (a) Picture of the electric field-driven strain measurement set up. The inside view of the electrode assembly is shown in (b, c). (b) shows the movable upper rod shaped (5 mm diameter) electrode. (c) shows the lower flat electrode. The MTI photonic displacement sensor is placed close to the top part of the upper electrode.

The strain measurement set up comprises of a flat bottom electrode and a spring loaded rod shaped (5 mm dia) movable upper electrode (Fig.1). Circular piezoceramic discs (typically of diameter 10- 12 mm) is placed in the centre of the flat bottom electrode and the movable spring loaded upper electrode gently presses the central region of the circular discs. In our set up, the upper moving electrode is kept at 0 V, and the flat bottom electrode is driven by a triangular bipolar/unipolar waveform of 1 Hz.

The thickness dependent on the field-driven strain as measured by the setup described above is shown in Figure 2. Similar to our previous reports [9,10], the strain increases considerably from 0.35% for 0.7 mm thick disc to 3% for 0.2 mm thick disc. Further, the bipolar strain curve becomes increasingly asymmetric with decreasing thickness, i.e., the strain decreases in the negative cycle and increases reamarkably in the positive cycle reaching > 3 %. Similar asymmetric bipolar strain has been reported by other groups [1-7] but without the recognition that it was a consequence of measurement on discs of small thicknesses on commercial set ups with similar electrode assembly as shown in Fig. 1.

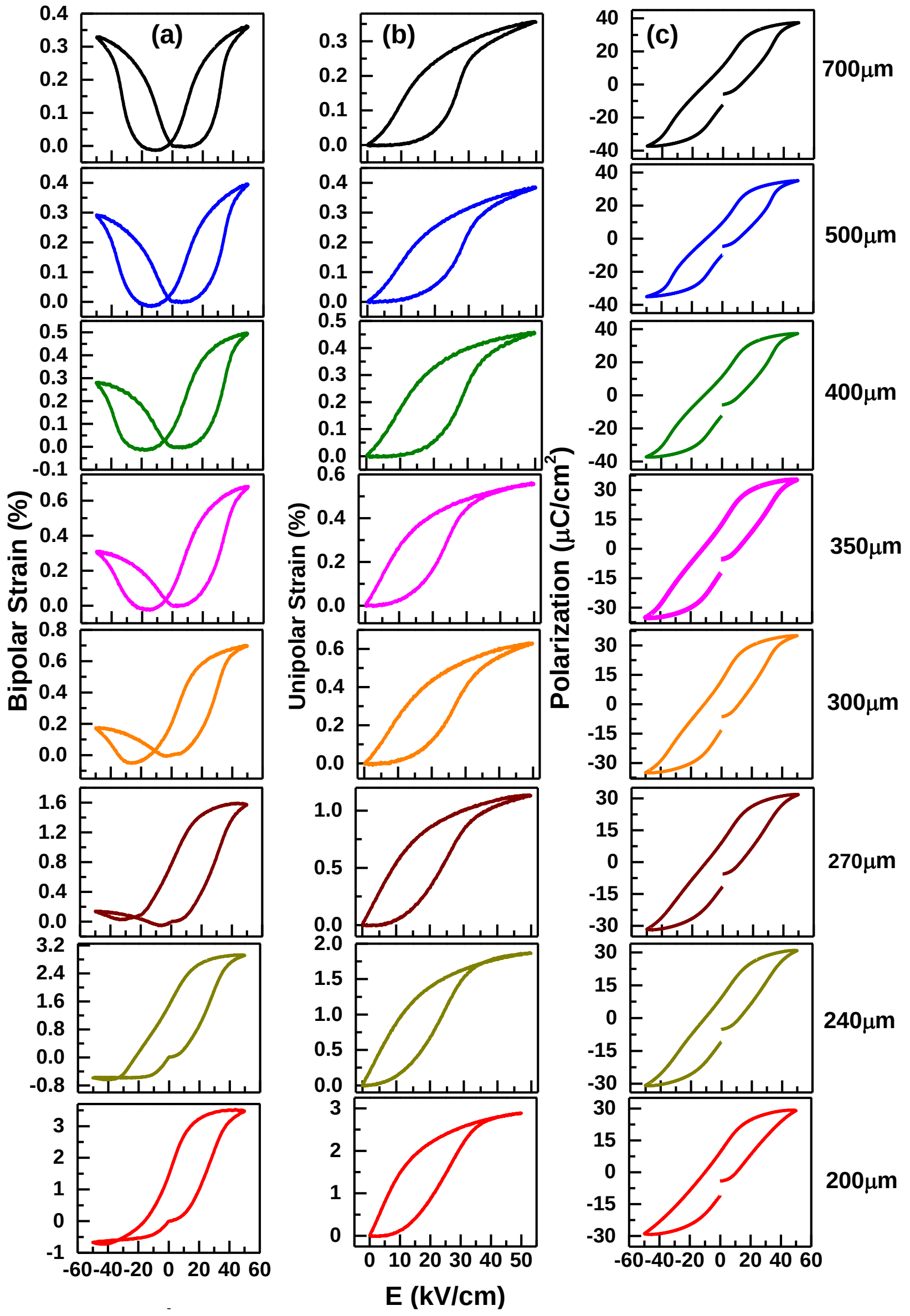

Fig. 2: (a) bipolar strain for different thickness of NBT-26ST. (b) unipolar strain of different thickness of NBT-26ST. (c) Bipolar polarization-electric field hysteresis loops.

In order to directly observe the nature of the deformation under field, an optical imaging system was coupled with a Critus Transmission Geometry measurement cell [12]. A rectangular section of the ceramic disc was cut with dimensions of 10 x 2 x 0.12 $mm^3$ with silver electrodes applied to the two opposing 10 x 2 $mm^2$ faces. Here, the sample is contacted between two point contacts within a silicon oil bath such that the sample may deform freely. A microscope and digital camera system was hardware synchronised with the applied electric field and the top contact displacement is also measured.

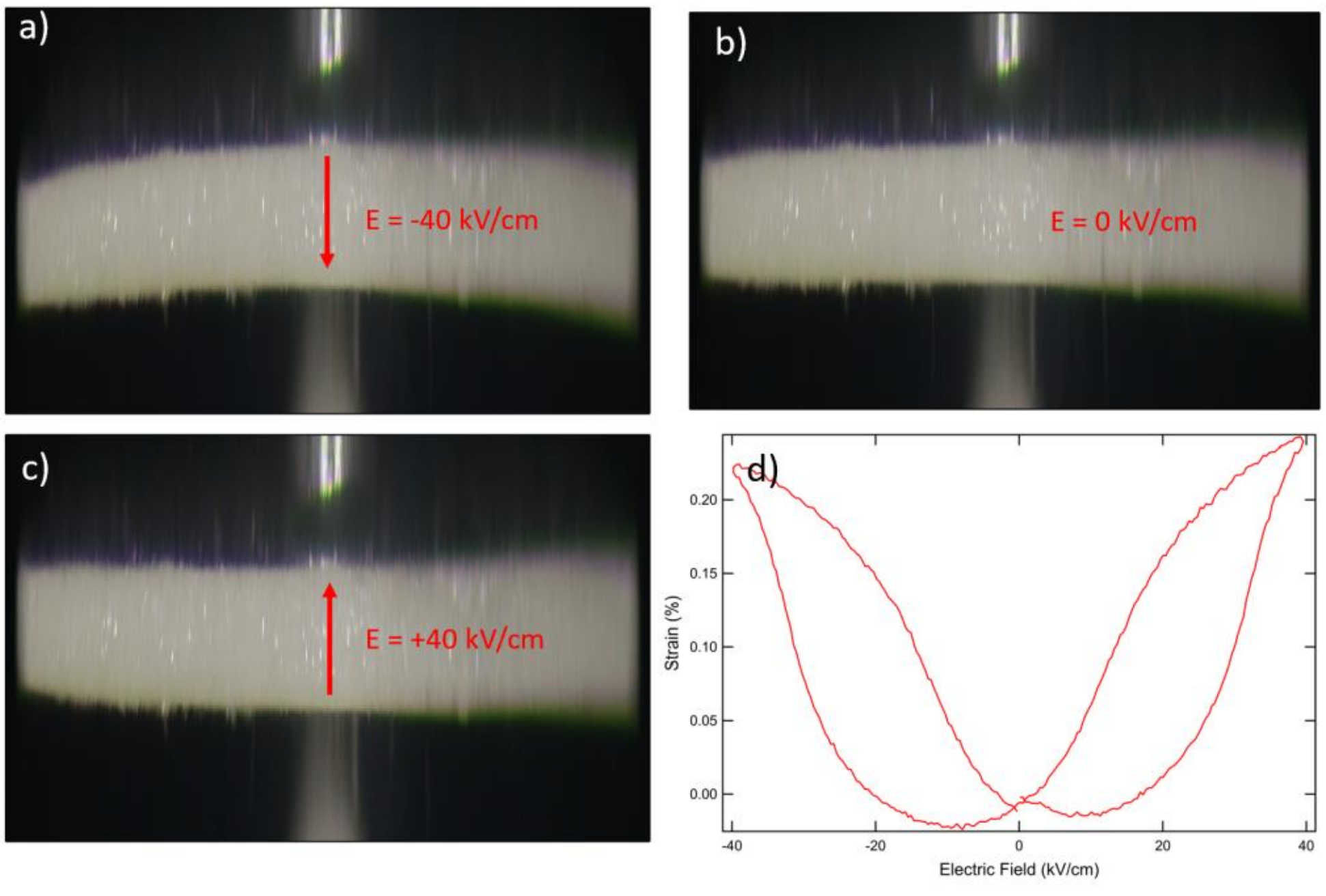

Fig. 3: Camera view of 0.1mm thin disc of NBT-26ST during application of electric field (a) - 50 kV/cm, (b) field switch off, and (c) + 50 kV/cm. Macroscopic strain measured by point contacts, measured simultaneously with the imaging. Note the images have been expanded vertically to visualise the response.

The images of the specimen at the maximum positive and maximum negative reveal the bending of the piezoceramic, Fig. 3a-c. The surface in contact with the positive potential acquires concave shape and the negative surface takes convex shape. Unlike Fig. 2, the bipolar strain measured with this set up is symmetric and ~ 0.2 %, Fig. 3d. This strain magntitude ~0.2 % is the real longitudinal electrostrain strain of the material and not the one seen in Fig. 2. As per the geometry of the electrodes assembly of the commercial strain measurement setup (Fig. 1), the photonic displacement sensor measures the lift of the upper electrode as the center of the disc bends up in the manner shown in Fig 3a and measures a large strain. When the field is reversed, the curvature of the bending reverses. In this situation, the central portion of the circular disc rests on the bottom flat electrode (Fig 1c) and the circumferential region lifts up. Since the upper electrode in our setup (Fig. 1) measures the displacement of only the central region of the disc, it measures almost no strain in the negative cycle.

Our experiments reveal that asymmetric strain loop with very large strain in only one half of bipolar field cycling seen in the thin discs (Fig.2) is therefore a consequence of two factors: (i) the specimen placed on a flat electrode and (ii) bending of the disc on application of electric field. The strain measured at the center of the disc by point contact method (Fig.3), on the other hand measures the true longitudinal strain. To understand the driving force for bending, we performed X-ray diffraction studies *in-situ* with an electric-field using a laboratory x-ray (CuK$\alpha_1$) diffractometer operating in reflection geometry. Field was applied across two gold-coated circular surfaces of the ceramic disc. X-ray diffraction pattern was recorded from the same surface by varying the magnitude and the sign of the field. Fig. 4 shows the x-ray Bragg profiles of pseudocubic $\{111\}_{pc}$ and $\{200\}_{pc}$ reflections of the virgin disc( at $E$ = 0, before applying the field), at $E$ = +50 kV/cm and after switching off the field on the thicker (0.7 mm) disc. Here, by positive field, we mean the diffracting surface has the positive voltage applied, and by negative field we mean the same diffracting surface has the negative voltage applied. Being close to the ergodic-non-ergodic relaxor boundary, the average structure of NBT-26ST is cubic and thus, each of the Bragg peaks are single and symmetric.

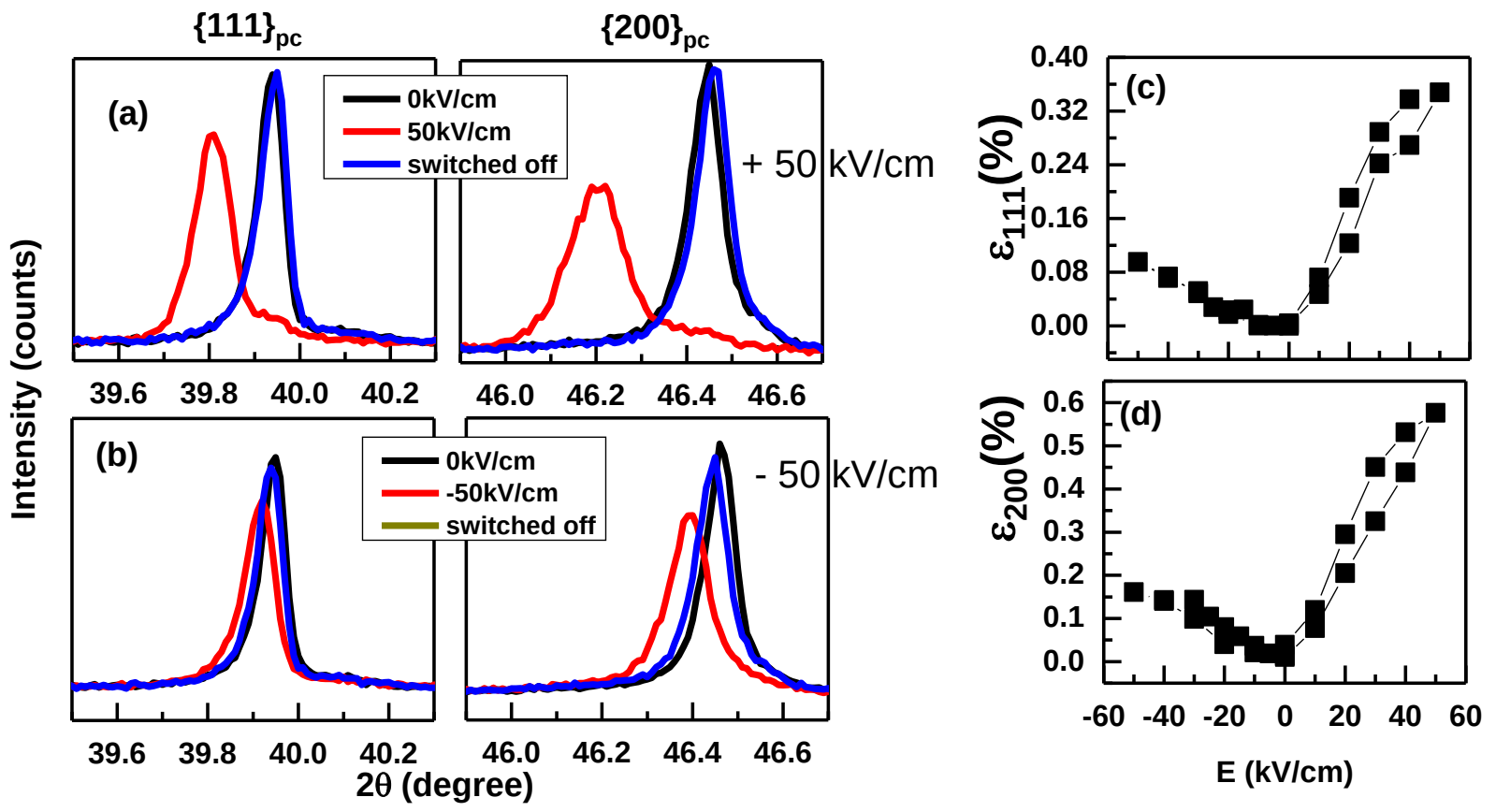

Fig. 4: The $\{111\}_{pc}$ and $\{200\}_{pc}$ Bragg peaks of x-ray diffraction (XRD) pattern of $Na_{0.5}Bi_{0.5}TiO_3$-0.26$SrTiO_3$ (NBT-26ST) in the application of (a)positive (0kV/cm, 50kV/cm and after switching off the field) and (b)negative (0kV/cm, -50kV/cm and after switching off the field) field cycles, (c, d) Electric field dependence of 111 lattice strain ($\varepsilon_{111}$ ) and 200 lattice strain of NBT-26ST under bipolar field cycling.

The effect of the electric field is to slightly broaden the profile and shift the peak positions to a lower 2θ value (Fig. 4a,b), indicating lattice expansion. It is interesting to note that the lattice expansion is significantly more for the positive field than the negative field. The lattice strain of the {111} and {200} planes as a function of the applied field was determined from the expression.

$$\varepsilon_{111}(E) = \{d_{111}(E) - d_{111}(E{=}0)\}/d_{111}(E{=}0) \quad \text{and} \quad \varepsilon_{200}(E) = \{d_{200}(E) - d_{200}(E{=}0)\}/d_{200}(E{=}0),$$

and plotted in Fig. 4c,d. The difference in the response of the grains near the positive and the negative surface of the disc implies that the positive and the negative surfaces will exhibit different deformation on the application of field. This difference will tend to bend the disc. The bending increases with decreasing thickness leading to increasing asymmetry in the bipolar strain when measured by set ups with electrode assembly like the one shown in Fig.1. In view of our study, the ultrahigh strain reported in the Pb-free piezoceramics [1-7] is in reality the measure of the lifting of the disc center by electrostrain measurement setups that include a flat surface in contact with the sample, similar to that shown in Fig. 1.

In summary, we have shown that the grains near the two surfaces of the $Na_{0.5}Bi_{0.5}TiO_3$ -based piezoceramic disc respond to electric fields differently and induces a propensity for the disc to bend. We propose that this asymmetry is brought about the migration of the oxygen vacancies in the presence of the field and their accumation near the negative surface. The phenomenon and the mechanism reported here is likely to be true for all materials exhibiting asymmetric bipolar strain loops when measured using commercial electrostrain measurement setups with the flat base electrode and small upper electrode. We hope our study unambiguously resolves the about the origin of the ultrahigh strain in Pb-free piezoceramics.

**Acknowledgments:** RR acknowledges the Science and Engineering Research Board for financial assistance (Grant Number: CRG/2021/000134).